\documentclass[sn-vancouver,Numbered]{sn-jnl}

\usepackage{graphicx}%
\usepackage{multirow}%
\usepackage{amsmath,amssymb,amsfonts}%
\usepackage{amsthm}%
\usepackage{mathrsfs}%
\usepackage[title]{appendix}%
\usepackage{xcolor}%
\usepackage{textcomp}%
\usepackage{manyfoot}%
\usepackage{booktabs}%
\usepackage{algorithm}%
\usepackage{bm}
\usepackage{dsfont}
\usepackage{algorithmicx}%
\usepackage{algpseudocode}%
\usepackage{listings}%
\usepackage[capitalize,noabbrev]{cleveref}

\begin{document}

\title[Group-wise normalization]{Group-wise normalization in differential abundance analysis of microbiome samples}


\author[1,*]{\fnm{Dylan} \sur{Clark-Boucher}}\email{dclarkboucher@fas.harvard.edu}

\author[1]{\fnm{Brent} \sur{Coull}}\email{bcoull@hsph.harvard.edu}

\author[2]{\fnm{Harrison T.} \sur{Reeder}}\email{hreeder@mgh.harvard.edu}

\author[3]{\fnm{Fenglei} \sur{Wang}}\email{fengleiwang@g.harvard.edu}

\author[3,4]{\fnm{Qi} \sur{Sun}}\email{qisun@hsph.harvard.edu}

\author[5,**]{\fnm{Jacqueline R.} \sur{Starr}}\email{spjst@channing.harvard.edu}

\author[1,3,4,**]{\fnm{Kyu Ha} \sur{Lee}}\email{klee@hsph.harvard.edu}

\affil[1]{\orgdiv{Department of Biostatistics}, \orgname{Harvard TH Chan School of Public Health}, \city{Boston}, \state{MA}, \country{United States}}
\affil[2]{\orgdiv{Biostatistics}, \orgname{Massachusetts General Hospital}, \city{Boston}, \state{MA}, \country{United States}}
\affil[3]{\orgdiv{Department of Nutrition}, \orgname{Harvard TH Chan School of Public Health}, \city{Boston}, \state{MA}, \country{United States}}
\affil[4]{\orgdiv{Department of Epidemiology}, \orgname{Harvard TH Chan School of Public Health}, \city{Boston}, \state{MA}, \country{United States}}
\affil[5]{\orgdiv{Channing Division of Network Medicine, Department of Medicine}, \orgname{Brigham and Women's Hospital}, \city{Boston}, \state{MA}, \country{United States}}
\affil[*]{Corresponding author}
\affil[**]{Co-senior authors}

\abstract{
\textbf{Motivation:}  A key challenge in differential abundance analysis of microbial samples is that the counts for each sample are compositional, resulting in biased comparisons of the absolute abundance across study groups. Normalization-based differential abundance analysis methods rely on external normalization factors that account for the compositionality by standardizing the  counts onto a common numerical scale. However, existing normalization methods have struggled at maintaining the false discovery rate in settings where the variance or compositional bias is large. This article proposes a novel framework for normalization that can reduce bias in differential abundance analysis by re-conceptualizing normalization as a group-level task. We present two normalization methods within the group-wise framework: group-wise relative log expression (G-RLE) and fold-truncated sum scaling (FTSS).
\newline
\textbf{Results:} G-RLE and FTSS achieve higher statistical power for identifying differentially abundant taxa than existing methods in model-based and synthetic data simulation settings, while maintaining the false discovery rate in challenging scenarios where existing methods suffer. The best results are obtained from using FTSS normalization with the differential abundance analysis method MetagenomeSeq.
\newline
\textbf{Availability and Implementation:} Code for implementing the methods and replicating the analysis can be found at our \textcolor{blue}{\href{https://github.com/dclarkboucher/microbiome_groupwise_normalization}{GitHub page}}. 
}
\maketitle

\section{Introduction}
\label{sec:intro}

Advancements in genetic sequencing technologies have enabled high-resolution cataloguing of the human microbiome, the microbial communities that collectively help modulate bodily processes in the mouth, skin, reproductive tract, gastrointestinal tract and other sites \citep{DeVos_Tilg_Van_Hul_Cani_2022, Grice_Segre_2012}. The role of the microbiome in human health has prompted research on finding specific microbes that differ in abundance according to site, exposures, disease status, or treatments, to name a few. In ``differential abundance analysis" (DAA), microbes are analyzed in categories defined by clustering the sequencing reads into phylogenetic taxa. 

One of the main challenges in DAA is that the microbial counts for each sample are typically compositional: across categories, the counts are constrained to sum to the total number of reads produced by the sequencing operation, referred to as that sample's ``library size" or ``sequencing depth" \citep{Tsilimigras_Fodor_2016, Gloor_Macklaim_Pawlowsky-Glahn_Egozcue_2017}. Thus, even when the goal is to compare the absolute abundance or quantity of microbes in the source, the data provide information only on the proportion of each taxon out of the whole ecosystem (the relative abundance). 
 To ignore this compositional structure and apply standard statistical methods would produce biased inference and inflated false discovery rates (FDRs), a phenomenon known as ``compositional bias" \citep{Gloor_Macklaim_Pawlowsky-Glahn_Egozcue_2017, Lin_Peddada_2020, Yang_Chen_2022}. 

Statistical methods for DAA that can account for compositional bias have generally fallen into two classes. The first class, which we call ``normalization-based methods," requires normalizing the taxon counts to account for their compositional nature. In these approaches, a ``normalization factor" is used to scale counts, enabling a meaningful comparison across samples. Normalization-based methods are used in common software packages such as edgeR \citep{Robinson_McCarthy_Smyth_2010}, DESeq2 \citep{Love_Huber_Anders_2014}, and MetagenomeSeq \citep{Paulson_Stine_Bravo_Pop_2013}. The normalization factor is computed externally using one of several available normalization methods \citep{Swift_Cresswell_Johnson_Stilianoudakis_Wei_2023}. The second class of DAA methods, which we call ``compositional data analysis methods," uses advanced statistical de-biasing procedures to correct model estimates without the need for external normalization \citep{Yang_Chen_2022}. Examples of analysis frameworks employing this approach include LinDA \citep{Zhou_He_Chen_Zhang_2022}, ANCOM-BC \citep{Lin_Peddada_2020}, and ALDEx2 \citep{Gloor_Macklaim_Pawlowsky-Glahn_Egozcue_2017}.

This article focuses on normalization-based methods, presenting a novel framework for the external calculation of normalization factors. Although the use of compositional data analysis methods has been increasing, normalization-based methods remain widely used in microbiome applications. Indeed, using Google Scholar to search for the term ``microbiome" in publications from January 1, 2021 to October 28, 2024 yielded 1,030 items citing the original MetagenomeSeq article, as well as 7,890 items citing the original DESeq2 article. This large body of work attests to the ongoing, frequent use of normalization-based methods for DAA of the microbiome. 

Several methods for calculating normalization factors have appeared in the literature. An early representative example is the relative log expression (RLE), which computes the normalization factor for a given sample by taking the across-taxa median of that sample's fold changes compared to each taxon's ``average" sample \citep{Anders_Huber_2010}. The key assumption of this method is that most samples should have similar true abundance to the average sample for most taxa; thus, a sample with high log fold changes should be counter-balanced with a high normalization factor. Many alternatives to RLE have been proposed following similar themes to account for different aspects of the data, such as zero-inflation, and have been reviewed in detail by others \citep{Lutz_Jiang_Neugent,Swift_Cresswell_Johnson_Stilianoudakis_Wei_2023}. However, these methods have generally suffered from poor FDR control and inflated type I error rates when used for DAA of microbiome data, especially when the differences in absolute abundance across study groups are large \citep{Zhou_He_Chen_Zhang_2022, Sohn_Du_An_2015}. 

We present a normalization framework that reduces bias in differential abundance analysis through its innovative use of group-level summary statistics of the subpopulations being compared. Several recent methods have been proposed in which compositional bias is quantified as a statistical parameter under different modeling assumptions, allowing for bias to be resolved by estimating this parameter \citep{Chen_Reeve_Zhang_Huang_Wang_Chen_2018, Kumar_Slud_Okrah_Hicks_Hannenhalli_}. In a similar fashion, we present a mathematical derivation that formally quantifies the statistical bias under an assumed model. We show that this derivation leads to a novel 
view of normalization as a group-level rather than a sample-level task. The group-wise view of normalization motivates two novel methods for calculating normalization factors: group-wise relative log expression (G-RLE), which applies RLE at the group level instead of the sample level, and fold truncated sum scaling (FTSS), which uses group-level summary statistics to find reference taxa. We compare the proposed methods to existing methods in extensive simulations that vary signal-to-noise ratio and signal sparsity. We also perform additional analyses of synthetic data based on two real-world microbiome datasets. Altogether, our results suggest the group-wise normalization framework offers higher power and more reliable FDR control than existing methods for normalization. 

\section{Methods}
\subsection{Group-wise normalization}
\label{sec:groupwise_norm}
\subsubsection{Compositional bias as a statistical parameter}
\label{sec:comp_bias}

In this section, we derive a formal characterization of the statistical bias from performing DAA on the observed taxon counts under a simple multinomial model. 
Suppose we have $n$ vectors of $q$ taxon counts $\bm{Y}_i = (Y_{i1},\dots,Y_{iq})^\text{T}$, $i\in\{1,\dots,n\}$, each representing a microbiome sample of the $i^\text{th}$ subject in a study. Define the library size for sample $i$  as $S_i=\sum_{j=1}^qY_{ij}$, and let $X_i$ be a binary covariate indicating group membership. Finally, let $\bm{A}_{i} = (A_{i1},\dots,A_{iq})^\text{T}$ denote the vector of the true absolute abundances corresponding to $Y_i$, and let $\bm{R}_i= (R_{i1},\dots,R_{iq})^\text{T}$ denote the true relative abundances corresponding to $\bm{A}_i$. Neither $\bm{A}_i$ nor $\bm{R}_i$ is observed.

We view the taxon counts as arising from the hierarchical data-generating mechanism
\begin{gather*}
    \log{A_{ij}} = \beta_{0j} + \beta_{1j}X_{i}\\
    \bm{R}_i = \frac{1}{\sum_{k=1}^qA_{ik}}\bm{A}_{i}\\
    \bm{Y}_i\mid (S_i,\bm{A}_i,X_i)\sim \text{Multinomial}(S_i,\bm{R}_i),
\end{gather*}
$i\in\{1,\dots,n\}$, $j\in\{1,\dots,q\}$. Under this model,  the absolute abundance ($A_{ij}$) is represented as a deterministic function of two parameters: $\beta_{0j}$, the absolute abundance in the samples with $X_i=0$, and $\beta_{1j}$, the log fold change in absolute abundance across covariate groups, the parameter of interest. Although, in reality, we do not believe the absolute abundance is equal for all samples within the same study group, this assumption provides mathematical convenience for developing a formal view of compositional bias. 

We consider fitting $q$ Poisson models of the form
\begin{gather*}
    \log (\lambda_{ij}) = \alpha_{0j} + \alpha_{1j}X_i + \log(S_i)\\
    Y_{ij}\sim \text{Poisson}(\lambda_{ij}),
\end{gather*}
for $j\in\{1,\dots,q\}$. While there are many other approaches for modeling sequencing data, such as a negative binomial regression, Poisson models are mathematically tractable due to the simplicity of the maximum likelihood estimator. Specifically, let 
\begin{gather*}
    \hat{R}^{(g)}_j = \frac{\sum_{i=1}^nY_{ij}\mathds{1}(X_i=g)}{\sum_{i=1}^nS_i\mathds{1}(X_i=g)}
\end{gather*}
denote the ``pooled" observed relative abundance of taxon $j$ in in group $g\in\{0,1\}$. Then maximum likelihood estimator of $\alpha_{1j}$ is $\hat{\alpha}_{1j} =\log (\hat{R}^{(1)}_j/\hat{R}^{(0)}_j)$, which we will call the \textit{observed} log fold change in contrast to $\beta_{1j}$, the true log fold change.

The statistical challenge with compositional data is that $\hat\alpha_{1j}$ is biased for estimating $\beta_{1j}$. Indeed, we show in the supplementary materials (Section 1) that
\begin{gather*}
    \hat{R}^{(g)}_j\overset{p}{\to}\frac{\exp(\beta_{0j}+\beta_{1j}g)}{\sum_{j=1}^q \exp(\beta_{0j}+\beta_{1j}g)}
\end{gather*}
as $n\to\infty$, where $\overset{p}{\to}$ denotes convergence in probability. Thus, 
\begin{align}
\label{eq:bias}
     \hat{\alpha}_{1j} \overset{p}{\to}  \beta_{1j} + \Delta,
\end{align}
where  $\Delta = \log \bigl( \sum_{j=1}^q\exp (\beta_{0j}) / \sum_{j=1}^q\exp (\beta_{0j} + \beta_{1j})\bigr)$ is an additive bias term that results from the compositional setting. We provide additional details for this derivation in Section 1 of the Supplementary Materials.

A critical feature of $\Delta$ is that it does not depend on the specific taxon. Indeed, $\Delta$ can be exactly interpreted as the log-ratio of the average total absolute abundance in the $X_i=0$ samples compared to the $X_i=1$ samples---a summary measure of the difference in microbial content across groups. Motivated by this observation, our proposed methodology contrasts from existing normalization in using its focus on group-level rather than sample-level summary statistics.

\subsubsection{Sample-wise versus group-wise normalization}
\label{sec:sample_vs_group}
Some commonly used normalization methods are RLE, cumulative sum scaling (CSS) \citep{Paulson_Stine_Bravo_Pop_2013}, trimmed mean of M-values (TMM) \citep{Robinson_Oshlack_2010}, geometric mean of pairwise ratios (GMPR) \citep{Chen_Reeve_Zhang_Huang_Wang_Chen_2018}, and Wrench \citep{Kumar_Slud_Okrah_Hicks_Hannenhalli_} (Table \ref{tab:norm_methods}). These methods involve the calculation of normalization factors, say $S_i^*$, $i\in\{1,\dots,n\}$, that reduce compositional bias when substituted for $S_i$ in the model's offset term. Using $S_i$ as in our derivation is called total sum scaling (TSS). The general approach is to divide each sample's observed taxon counts or relative abundances by the corresponding values of some reference---which could be a single, pre-specified sample or a pooled version of the full dataset---then set $S_i^*$ as a summary statistic of these ratios. Ideally, under sparsity assumptions about the proportion of taxa that are differentially abundant, the normalization factors are proportional to the unknown sampling fractions, enabling inference on the absolute abundance scale \citep{Lin_Peddada_review_2020}.

\begin{table}[h]
\caption{Normalization methods}\label{tab:norm_methods}
\footnotesize
\renewcommand{\arraystretch}{1.8}
\begin{tabular*}{\textwidth}{@{\extracolsep\fill}p{1.45cm}p{4cm}p{6cm}}
\toprule%

Method & Software & Normalization Factor \\
\midrule
TSS   & N/A & Library size\\
CSS \citep{Paulson_Stine_Bravo_Pop_2013} & \texttt{metagenomeSeq} R package  & Truncated library size to exclude outliers\\
RLE \citep{Anders_Huber_2010} &  \texttt{edgeR} R package & Median of count ratios compared to the average sample\\
GMPR  \citep{Chen_Reeve_Zhang_Huang_Wang_Chen_2018} & \texttt{GMPR} package on GitHub  & Robust average of sample-to-sample comparisons to account for zero-inflation.\\
TMM  \citep{Robinson_Oshlack_2010} &  \texttt{edgeR} R package   & Trimmed and weighted average of fold changes compared to a reference sample\\
Wrench  \citep{Kumar_Slud_Okrah_Hicks_Hannenhalli_} & \texttt{Wrench} R package  & Robust average of model-regularized log fold changes\\

\bottomrule
\end{tabular*}

\end{table}

The idea of the proposed normalization framework is to shift the focus from sample-to-sample comparisons to group-level comparisons. Specifically, for RLE, GMPR, TMM, and Wrench, the summary statistic underlying the normalization factor is computed at the sample level, summarizing the fold changes between that sample and the ``typical" sample. However, as elucidated by $\Delta$ in Equation \ref{eq:bias}, the compositional estimation bias reflects a difference at the level of the \emph{group}: the ratio of the average total absolute abundance between study groups. The group-wise framework makes use of this observation by relying on summary statistics of the log fold changes between groups rather than samples---re-conceptualizing normalization as an intrinsically group-level task . Next, we present two new normalization methods that use this framework.


\subsubsection{G-RLE: RLE normalization at the group level}
\label{sec:grle}
The first proposed method is group-wise relative log expression (G-RLE): a modified version of RLE that depends on group-level instead of sample-level fold changes. Specifically, let $\tilde{R}_j = \sqrt{\hat{R}_j^{(0)}\hat{R}_j^{(1)}}$ be the geometric mean of the group-level relative abundances, $j\in\{1,\dots,q\}$. Now, for group $g\in\{0,1\}$, let $C^{(g)} = \text{median}_j (\hat{R}^{(g)}_j / \tilde{R}_j)$. The normalization factors are calculated as $S^\text{G-RLE}_i = S_i \times C^{(X_i)}$.

In this definition, the group-level relative abundances $(\hat{R}_1^{(g)},\hat{R}_2^{(g)},\dots,\hat{R}_q^{(g)})$, $g\in\{0,1\}$, are effectively treated as two microbiome samples, and the summary statistics ($C^{(g)}$) are computed as if the dataset contained only those samples. The normalization factors are then computed for each individual by their corresponding group-level statistic. The final normalization factors, $S_i^\text{G-RLE}$, remain sample-specific, and their magnitudes depend on the summarized data across treatment groups. If the median true log fold change is zero, using $S^\text{G-RLE}_i$ as the offset eliminates  $\Delta$ from Equation \ref{eq:bias} (Supplementary Material, Section 1).

\subsubsection{FTSS: Scaling normalization based on reference taxa}
\label{sec:FTSS}
The second proposed method is fold-truncated sum scaling (FTSS): a truncation-based normalization approach that uses ``reference taxa" that are assumed to be equally abundant under the considered model. Identifying reference taxa is a common strategy in both normalization-based and compositional data analysis methods for DAA \citep{Yang_Chen_2022}. The notion of reference taxa has direct relevance to the compositional bias  from Equation  \ref{eq:bias}; indeed, if $S_i$ were restricted to exclude taxa that are differentially abundant, the bias term $\Delta$ in would disappear (Supplementary Material, Section 1).

FTSS builds on the convergence established in Equation \ref{eq:bias}, which implies that, in large samples, the log fold changes in the observed relative abundance ($\hat\alpha_{1j}$) should approximate the true log fold changes of interest ($\beta_j$) up to the additive constant $\Delta$. We now make the additional assumption that only a minority of taxa are differentially abundant; thus, most of the $\beta_j$s should be zero, and their corresponding $\hat\alpha_{1j}$s should be near $\Delta$. 

In FTSS, the compositional bias is estimated as $\hat\Delta =\text{Mode}_j\bigl(\hat{\alpha}_{1j}\bigr)$ using Gaussian kernel density estimation, and the library size is restricted so that it includes only taxa for which $\hat{\alpha}_{1j}$ is within a pre-specified percentile of $\hat\Delta$. Specifically, let $\rho(x)$ be a function denoting the proportion of taxa whose $\hat\alpha_{1j}$ is at most $x$, and let $p^*$ be the pre-specified proportion of taxa to be included in the truncated sum. Then
\begin{gather*}
    S_i^\text{FTSS} = \sum_{j=1}^qY_{ij} \mathds{1}\biggl(\rho(\alpha_{1j})\in 
    \bigl(\rho(\hat\Delta) - p^*/2, \rho(\hat\Delta) + p^*/2\bigr)\biggr).
\end{gather*}
The logic of the procedure is that taxa whose observed log fold changes are close to $\Delta$ should have true log fold changes equal to (or close to) zero; thus, taxa with observed log fold changes near $\hat\Delta$ should form a reasonable reference set. 

\begin{figure*}[!t]
\makebox[\textwidth][c]{
\includegraphics[width=7.5in]{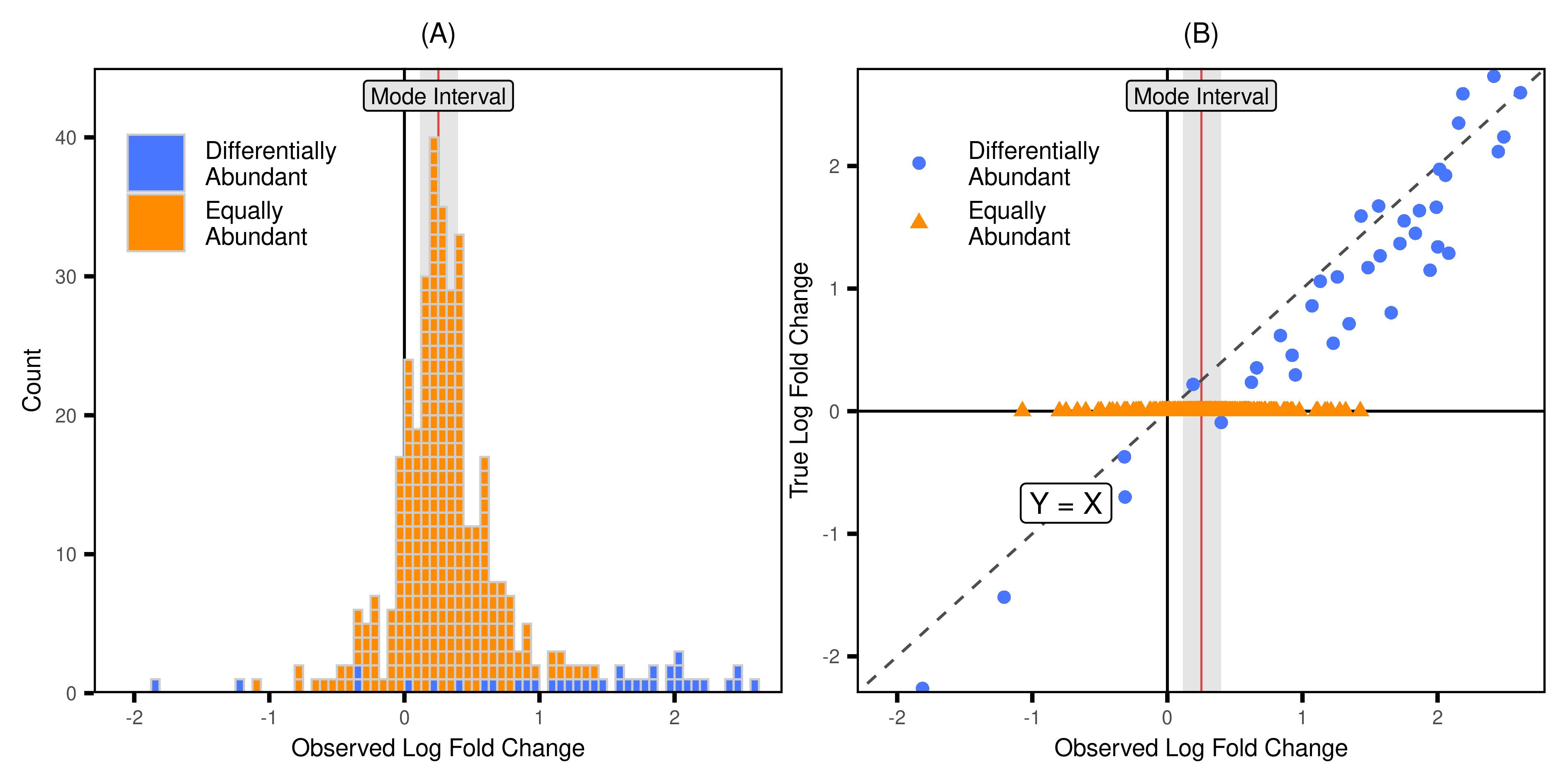}}
\caption{\textbf{Identification of reference taxa in FTSS}. (A) Stacked histograms of the observed log fold changes from a synthetic microbiome dataset based on the MLVS/MBS data. The orange histogram shows equally abundant taxa and blue shows differentially abundant taxa.  (B) A plot of the observed versus true log fold changes. The gray band shows a 40\% interval around the estimated mode from the distribution shown in (A). Taxa that fall within the gray band form the truncated scaling factor in FTSS normalization.}
\label{fig:log_fc}
\end{figure*}

We used a synthetic microbiome dataset to illustrate how FTSS uses the observed log fold changes. The true non-zero log fold changes were generated from a Normal(1, 1) distribution for only 10\% of the taxa. The observed log fold changes of the equally abundant taxa form a tight, mound-shaped distribution that is off-center from zero (Fig. \ref{fig:log_fc} (A)).  This clustering suggests that the reference taxa can be chosen based on the proximity of their log fold changes to the mode log fold change. When the observed log fold changes are compared to the true log fold changes (Fig. \ref{fig:log_fc} (B)), the trend in points lies directly below the line Y = X, suggesting a constant bias term as characterized by $\Delta$ in Equation \ref{eq:bias}. The gray ribbon covers taxa that fall within a $p^*=40\%$ percentile interval around the mode log fold change; such taxa form the truncated library size. Moreover, although a differentially abundant taxon is mistakenly included in the reference set, this is unlikely to introduce bias since the true log fold change of that taxon is small.  

\subsection{Model-based simulations}
\label{sec:modelbased}

We compared the performance of the proposed FTSS and G-RLE normalization methods to that of TSS, TMM, RLE, GMPR, CSS, and Wrench in simulations using the DAA methods edgeR \citep{Robinson_McCarthy_Smyth_2010}, DESeq2 \citep{Love_Huber_Anders_2014}, and metagenomeSeq \citep{Paulson_Stine_Bravo_Pop_2013} (summarized in Supplementary Material, Section 2). Every combination of normalization method and DAA method was evaluated by two metrics: the true positive rate (TPR) for detecting $\beta_{1j}$ at fixed values of the false positive rate (FPR), and the TPR and false discovery rate (FDR) at a 0.05 nominal FDR after applying the Benjamini and Hochberg procedure \citep{Benjamini_Hochberg_1995}. TPR is defined as the average proportion of differentially abundant that are detected, whereas FPR is the average proportion of equally abundant that are detected. We averaged these values over 1,000 replications in each of 18 simulation settings.

\subsubsection{Data generation and settings}
We generated data from a hierarchical multinomial model that incorporated subject-level variance and zero-inflation.  Several methods for simulating microbiome data have appeared in the literature; our approach was based on the models presented by Lee et al. (2020) \citep{Lee_Coull_Moscicki_Paster_Starr_2020} and Chiquet et al. (2021) \citep{Chiquet_Mariadassou_Robin_2021} to induce correlations among the positive- and zero-valued taxon counts. Specifically, for $i\in\{1,\dots,n\}$, we generated initial, positive absolute abundances $\bm{A}_i = (A_{i1},\dots,A_{iq})^\text{T}$ of $q$ taxa under the model $\log(A_{ij}) = \beta_{0j} + \beta_{1j}X_i + V_{ij}$, where the $V_{ij}$'s are correlated error terms. $\bm{V}_i = (V_{i1},\dots,V_{iq})^\text{T}$ was sampled from a multivariate normal distribution, MVN($\bm{0}$, $\sigma^2\Sigma_v$), where $\sigma^2$ is the variance and $\Sigma_v$ is a correlation matrix.  

 We incorporated zero-inflation with a multivariate probit distribution \citep{Ashford_Sowden_1970} by introducing latent variables, $\bm{Z}_i = (Z_{i1},\dots,Z_{iq})^\text{T}\sim \text{MVN}(\bm{0},\Sigma_z)$, where the correlation matrix $\Sigma_z$ induces relationships in the presence and absence of taxa in each sample. Then $B_{ij}$, the final absolute abundance of taxon $j$ in sample $i$, and $B^\prime_{ij}$, the corresponding relative abundance, are set to be  
\begin{gather*}
    B_{ij} = A_{ij}\times\mathds{1}(Z_{ij} > \alpha),\text{     }j\in\{1,\dots,q\},\\
    B^\prime_{ij} = \frac{B_{ij}}{\sum_{k=1}^qB_{ik}},\text{     }j\in\{1,\dots,q\},
\end{gather*}
where the parameter $\alpha$ controls the degree of zero inflation. Finally, we sampled the library sizes $S_i$ from a negative binomial distribution with mean  $\mu_S$ and variance $\gamma^2_S$ and generated the final taxon counts as $\bm{Y}_i \sim \text{Multinomial}(S_i, \bm{B}_i^\prime)$.

In these simulations, the number of taxa ($q$) was fixed at 300, and the parameter $\alpha$ was chosen so that the marginal probability of each $B_{ij}$ being zero was 0.50. The $\beta_{01},\dots,\beta_{0q}$ values were sampled from a $\text{Normal}(0,1)$ distribution. The correlation matrix $\Sigma_v=\Sigma_z$ was sampled from a Wishart distribution with $2q$ degrees of freedom and identity scale matrix. We set $\mu_S=60000$ and $\gamma_S=150000$.

The following parameters were varied: $\beta_{1q},\dots,\beta_{1q}$, which were defined by sampling the first 10\%, 20\%, or 30\% of the entries from a $\text{Normal}(1,1)$ and setting the others to zero;  $\sigma^2$, which was set to 1, 1.5, or 2 as ``low", ``medium", and ``high" variance settings, respectively; and $n$, which was set to 200 or 500. 
Half the data were given $X_i=0$ and half $X_i=1$. The combinations of $\bm\beta_1$, $\sigma^2$, and $n$ produced 18 different settings in total. 
We also evaluated the method's performance in a more complex regression setting that includes adjustment for a confounding variable (described and discussed in Supplementary Material, Section 3).

\subsection{Synthetic data simulations}
\label{sec:Synthetic}
Microbiome sequence data exhibit complicated correlation structures that are hard to replicate in simulations. Thus, we performed additional simulations based on synthetic data derived from two real-world microbiome datasets. The first dataset is from the Pediatric HIV/AIDS Cohort Study (PHACS), and the second dataset combines data from the Men's Lifestyle Validation Study (MLVS) and the Mind-Body Study (MBS). We refer to the two datasets as simply ``PHACS" or ``MBS/MLVS", respectively.

\subsubsection{Datasets}
The Adolescent Master Protocol is a prospective cohort study conducted by the PHACS network on the health effects of HIV infection among perinatally HIV-exposed youth. From September 2012 to January 2014, subgingival dental plaque samples were collected on 254 subjects ages 10-22 years \citep{VanDyke_Patel_Siberry_Burchett_Spector_Chernoff_Read_Mofenson_Seage_2011,Shiboski_Yao_Russell_Ryder_VanDyke_Seage_Moscicki_2018, Schulte_King_Paster_Moscicki_Yao_VanDyke}. Participants were excluded if they had antibiotic exposure in the prior 3 months. DNA was isolated from plaque specimens and 16S rDNA sequenced \citep{Caporaso_Lauber_Walters_Berg-Lyons_Lozupone_Turnbaugh_Fierer_Knight_2011,Gomes_Berber_Kokaras_Chen_Paster_2015}. The sequencing reads were trimmed, filtered, and grouped using the DADA2 pipeline, and reads matched to the curated Human Oral Microbiome Database (99.9\% of reads matched to the species or genus level) \citep{Dewhirst_Chen_Izard_Paster_Tanner_Yu_Lakshmanan_Wade_2010}. The final dataset included taxon counts for 344 OTUs on each of 254 samples.

MLVS was a one-year follow-up study of the Health Professionals Follow-Up Study (HPFS), an ongoing prospective cohort study that began in 1986 \citep{Li_Wang_Li_Ivey,Li_Wang_Satija_Ivey,Mehta_Abu-Ali_Drew_Lloyd-Price_Subramanian_Lochhead_Joshi_Ivey_Khalili_Brown}. HPFS comprised 51,529 US male health professionals; of these, 307 participants were included in the MLVS and provided up to two pairs of self-collected stool samples from 2011 to 2013. 

MBS was a one-year follow-up study of the Nurses' Health Study II (NHSII), an ongoing prospective cohort study of female registered nurses that began in 1989 \citep{Huang_Trudel-Fitzgerald_Poole_Sawyer_Kubzansky_Hankinson_Okereke_Tworoger_2019, Ke_Guimond_Tworoger_Huang_Chan_Liu_Kubzansky_2023}. From 2013 to 2014, 213 of the 116,429 NHSII participants were enrolled into MBS and had their gut microbiome sequenced based on two self-collected stool samples. For both MLVS and MBS, participants were free of coronary heart disease, stroke, cancer, and major neurological disease at the time of enrollment. 

Sequencing procedures for MBS and MLVS were identical. Reads were processed through KneaData 0.3. High quality reads were taxonomically profiled using MetaPhlAn 4 \citep{Blanco-Míguez_Beghini_Cumbo_McIver_Thompson_Zolfo_Manghi_Dubois_Huang_Thomas_etal._2023}, resulting in 2201 species-levels taxa in the HPFS cohort and 1860 in the NHSII cohort. We kept taxa present in at least 10\% of samples with a minimum relative abundance of 0.0001 and present in both cohorts. The initial number of taxa was 1,860 in the MBS and 2,201 in the MLVS; after filtering, 372 taxa remained. Following the approach used in other studies from the same cohorts, we combined the MLVS and MBS data into a single dataset \citep{Wang_Glenn_Tessier_Mei_Haslam_Guasch-Ferré_Tobias_Eliassen_Manson_Clish_etal._2024, Li_Hur_Cao_Song_Smith-Warner_Liang_Mukamal_Rimm_Giovannucci_2023}. The combined data included 213 female and 307 male participants.

\subsubsection{Data generation and settings}
With either dataset, we perturbed the taxon counts to induce signals corresponding to a binary covariate while preserving the overall structure, i.e., while ensuring that the variance, correlations, zero-inflation and relative abundance of the synthetic data mirrors those of the real data. 

We generated the synthetic data with a re-sampling scheme similar to down-sampling \citep{Pereira_Wallroth_Jonsson_Kristiansson_2018}. Specifically, we began with the observed taxon count matrix $\bm{Y}$ from either PHACS or MLVS/MBS, with $n$ observations and $q$ OTUs. Let $\bm{\beta_1}$ be a vector of $q$ log fold changes. To generate a single synthetic dataset, we split $\bm{Y}$ in half at random to define $\bm{Y}^{(0)}$, the subjects with $X_i=0$, and $\bm{Y}^{(1)}$, the subjects with $X_i=1$. While $\bm{Y}^{(0)}$ remained unchanged, for each $i\in\{1,\dots,n/2\}$, we re-sampled the taxon counts of the $i^\text{th}$ subject in $\bm{Y}^{(1)}$ as $\bm{\tilde{Y}}^{(1)}_i \sim \text{Multinomial}(S_i,\bm{p}_i)$, where $S_i$ is the library size and $\bm{p}_i=(p_{i1},\dots,p_{iq})$ is a vector of probabilities such that $p_{ij}\propto Y^{(1)}_{ij}\exp(\beta_{1j})$. We generated $\bm{\beta_1}$ in the same fashion as for the model-based settings described above, resulting in settings with 10\%, 20\%, or 30\% signals.

For analysis, we used the same DAA methods, normalization methods, and evaluation metrics as for the model-based simulations. We evaluated methods' performance in 1,000 replications of three settings,

\section{Results}\label{results}
\subsection{Simulation results}\label{results1}
\begin{figure*}[t]
\makebox[\textwidth][c]{
\includegraphics[width=7.5in]{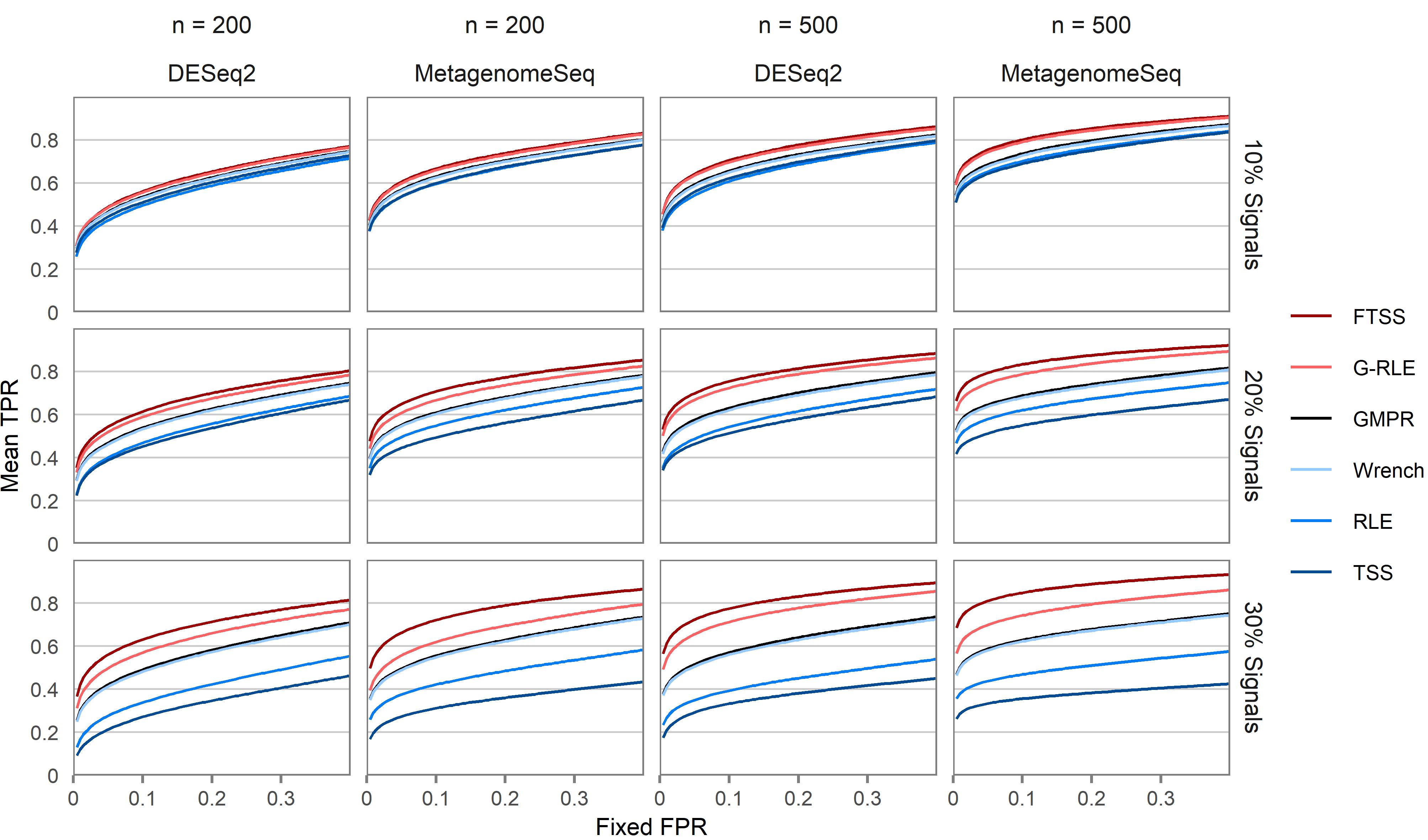}}
\caption{\textbf{True positive rate for detecting differentially abundant taxa for simulations with high variance}. Value shown is the mean over 1,000 replicates of each setting. Columns indicate sample size and DAA method while rows indicate the proportion of differentially abundant taxa out of 300.}
\label{fig:sim_tpr}

\end{figure*}

For datasets with a ``high" variance, 200 or 500 samples, and in which either DESeq2 or MetagenomeSeq were used for DAA, G-RLE and FTSS outperformed existing normalization methods at every value of the FPR (Figure \ref{fig:sim_tpr}). This improvement in TPR was slight when the signal percentage was 10\% but increased to 10-15\% higher than the best-performing existing method, GMPR, when the signal percentage was raised to 20\% or 30\%. Results for the low and medium variance simulations were similar (Supplementary Material, Section 4).

\begin{figure*}[b]
\makebox[\textwidth][c]{
\includegraphics[width=7.5in]{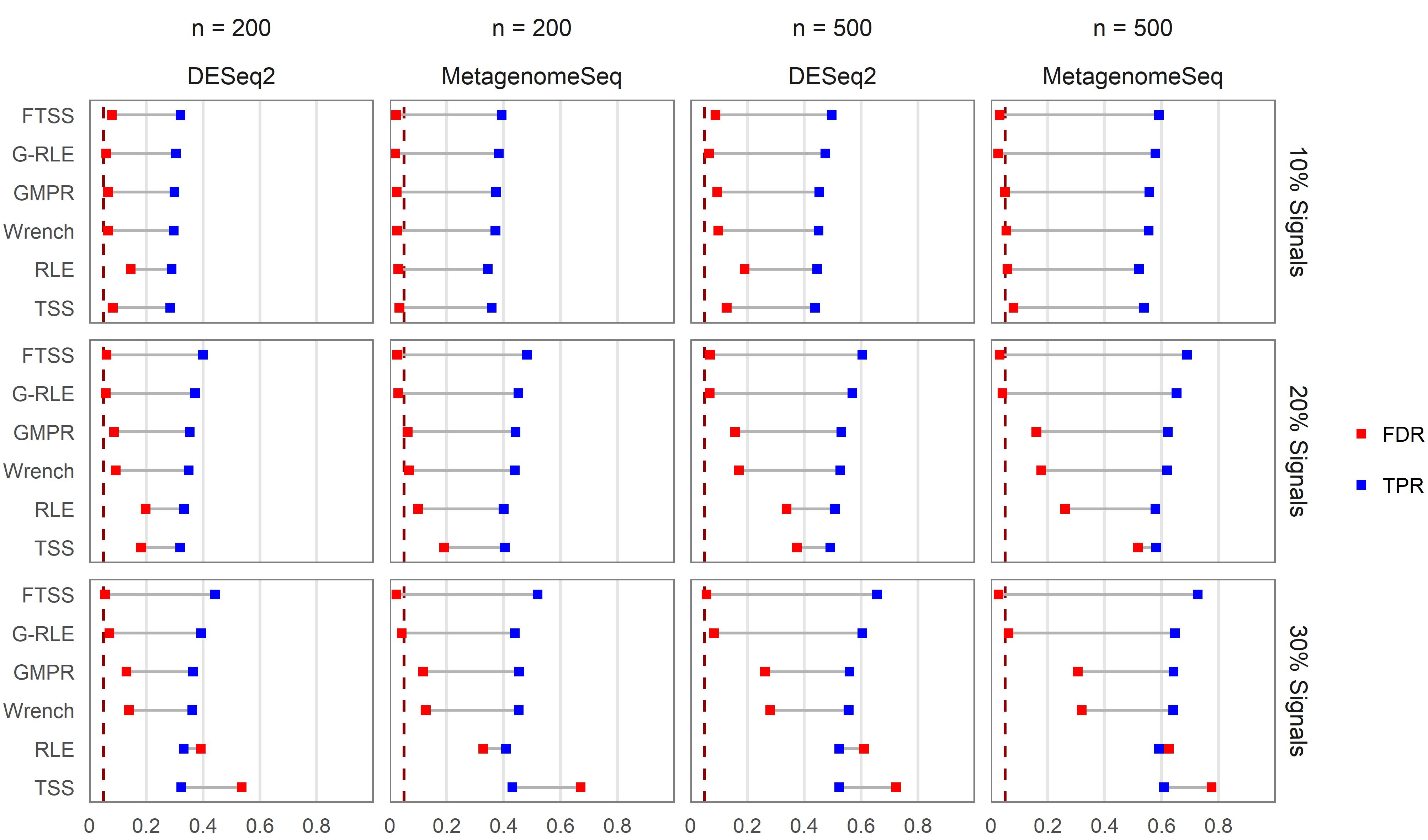}}
\caption{\textbf{Performance of normalization methods at nominal false discovery rate for simulations with high variance}. The true positive rate (blue) and observed false discovery rate (red) corresponding to a nominal false discovery rate of 0.05 (the dashed line). Value shown is the mean over 1,000 replicates of each setting. Columns indicate sample size and DAA method. Rows indicate the proportion of differentially abundant taxa out of 300 total.}
\label{fig:sim_mix}
\end{figure*}

At a nominal FDR of 0.05  (Figure \ref{fig:sim_mix}), FTSS attained the highest TPR in every setting and G-RLE the second-highest in most settings, with FTSS often beating GMPR by more than 5\%. Meanwhile, both FTSS and G-RLE either maintained the nominal FDR or close to it in settings where existing methods struggled, specifically, with 30\% signals and a small sample size, or 20-30\% signals and a larger sample size. In one of the more extreme cases---using MetagenomeSeq on datasets with 30\% signals, high variance, and $n=500$---FTSS achieved around 73\% TPR and 3\% FDR, whereas GMPR had only 64\% TPR and 31\% FDR. Results for the low and medium variance settings were similar and had better FDR control across the board (Supplementary Material, Section 4). Finally, our additional results suggest the group-wise framework remains effective and outperforms existing methods in analyses adjusted for a confounding variable (Supplementary Materials Section 3).

\subsection{Synthetic Data Results}
\label{sec:synth_results}
\begin{figure*}[htbp]

\makebox[\textwidth][c]{
\includegraphics[width=7.5in]{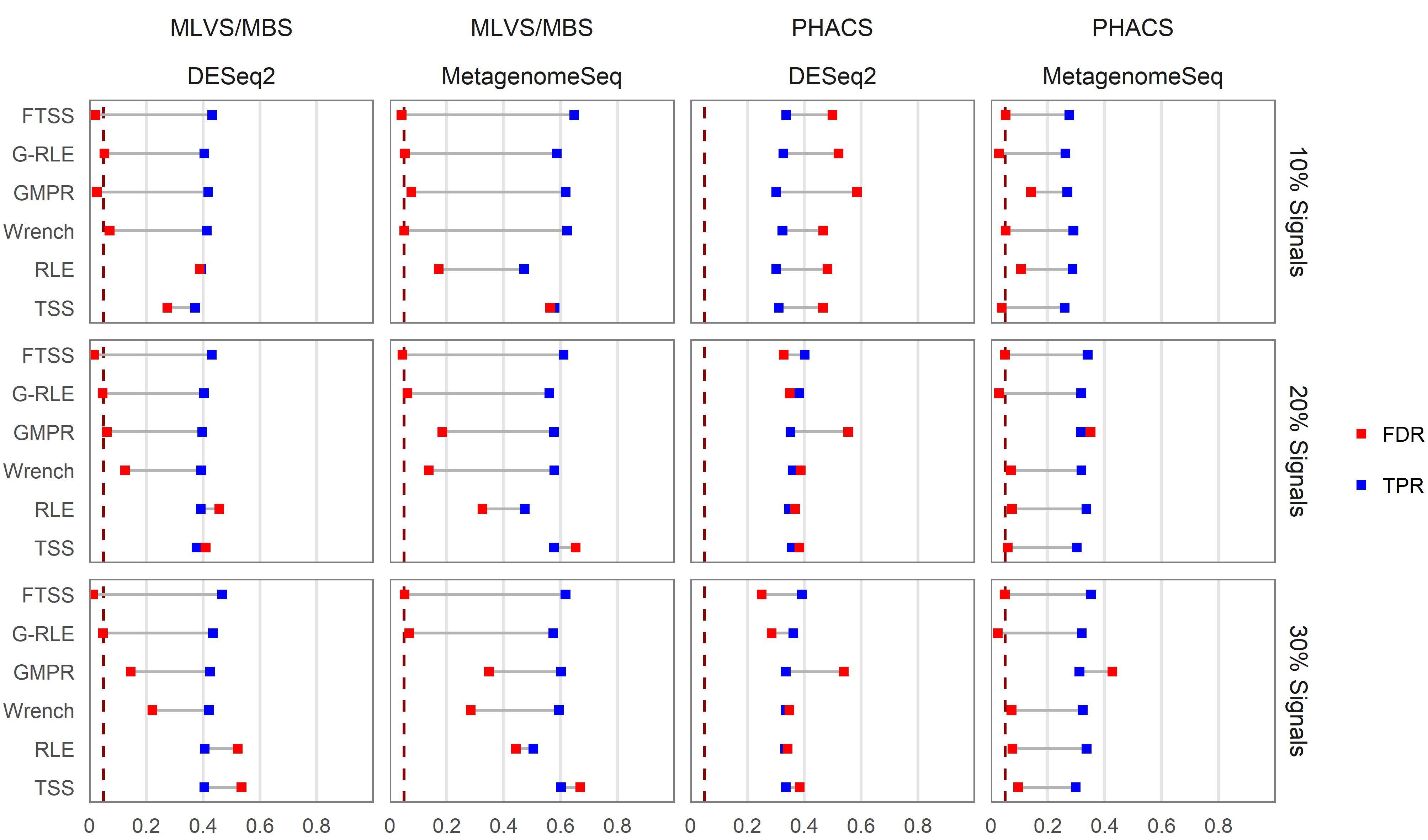}}
\caption{\textbf{Performance of normalization methods at nominal false discovery rate in synthetic data simulations}. The true positive rate (blue) and observed false discovery rate (red) corresponding to a nominal false discovery rate of 0.05 using either the MLVS/MBS or PHACS dataset. Value shown is the mean over 1,000 replicates of each setting. Rows indicate the proportion of differentially abundant taxa. }
\label{fig:synth_mix}
\end{figure*}

Analysis of synthetic data based on the MLVS/MBS  data largely matched those from the model-based simulations, with both DESeq2 and MetagenomeSeq maintaining the FDR or close to it when paired with either FTSS or G-RLE (Figure  \ref{fig:synth_mix}). In contrast, the competing normalization methods struggled in their FDR control when 20\% or 30\% of the taxa were differentially abundant. FTSS combined with MetagenomeSeq maintained the FDR near 0.05 and achieved the highest power in every setting. 

Analysis of synthetic data based on the PHACS gave different results depending on the choice of DAA method. With MetagenomeSeq, comparisons across normalization methods gave similar results to those described above for the model-based simulations and the other synthetic dataset.  In contrast, with DESeq2, no normalization method controlled the FDR, which often exceeded the TPR. 

\section{Discussion}

In this work, we presented a novel framework for normalization that exploits assumptions about group-level log fold changes, using normalization factors that directly target the compositional bias term across comparison groups. In numerical studies, the proposed G-RLE and FTSS normalization methods demonstrated substantial improvements in sensitivity and FDR control over existing techniques, especially in challenging data scenarios characterized by a high proportion of differentially abundant taxa or high noise. Between the two newly proposed methods, FTSS was more consistent in maintaining the FDR and achieved slightly higher power compared with G-RLE.  

The performance gap between the group-wise normalization method and existing methods may be rooted in their fundamental difference in approach. While the approach of most existing normalization methods is to use the sample-level log fold changes to estimate sampling fractions, the group-wise framework uses comparisons of the pooled group-level data to estimate a specific mathematical bias term. Thus, it is conceivable from a statistical viewpoint that estimating one parameter---the bias term---is an easier task than estimating $n$ sampling fractions based on potentially noisy comparisons of individual samples. This benefit may be especially advantageous in the presence of zero-inflation: The group-level pooled data are strictly positive. Thus, the summary statistics used by the group-wise framework may have an inherent robustness to zero-inflated data, in contrast to existing approaches, which may struggle to estimate the sampling fractions of samples with excessive zeros.

The foundations of the group-wise framework bear similarity to those of Wrench. Like our own methods, Wrench treats compositional bias as an estimable parameter under a hierarchical multinomial model \citep{Kumar_Slud_Okrah_Hicks_Hannenhalli_}. Wrench estimates this parameter by fitting a mixed effects model that applies shrinkage to the sample-level log fold changes compared to the average sample, while incorporating a hurdle component to model structural zeros. Wrench's relatively poor performance compared to G-RLE and FTSS may be attributable to its assumption that the true log fold changes have a mean of zero, a condition that did not align with our practical simulation designs. In contrast, G-RLE operates under the assumption that the median log fold change is zero, and FTSS assumes only that the most common log fold change is zero, providing greater robustness to the  distribution of the unknown signals. However, Wrench may still enjoy strong performance in real-world settings that more closely match its modeling assumptions. 

Our simulations revealed a scenario in which all normalization methods, including the proposed frameworks, failed to control the FDR---namely, when edgeR or DESeq2 was applied to the PHACS data. This was likely due to challenges such as extremely high sparsity, as approximately 40\% of taxa had fewer than 5 counts per sample on average. We obtained stronger FDR control on the PHACS data when pairing the group-wise gramework with MetagenomeSeq, a tool specifically designed for zero-inflated datasets.

We also note that, like Wrench, the proposed group-wise framework is designed explicitly for settings in which the main covariate of interest is categorical, not continuous. Consequently, we view the group-wise normalization framework as primarily suited for DAA comparing study groups, exposure statuses, or subpopulations, which is a central task in microbiome data analysis. A key direction for further investigation is the the development of methods to calculate normalization factors that can reduce compositional bias in more complex settings not considered in this study, especially those in which the main covariate of interest is continuous. 

In conclusion, the proposed normalization framework enables more robust statistical analysis for investigating the association between the covariate of interest and taxa levels than existing tools. The mathematical derivation, based on the data-generating mechanism developed in this work, along with the publicly available software and findings from our simulation studies, will support a broad range of future research on the microbiome.

\section{Competing interests}
The authors have declared there are no competing interest

\section{Author contributions statement}

D.C., K.L., and H.T.R. developed the methodology; D.C. performed the analysis; D.C., J.R.S and K.L. conceptualized the project; K.L., J.R.S., and B.A.C. supervised the project; F.W. and Q.S. were responsible for data acquisition and cleaning; D.C. wrote the initial draft of the manuscript; all authors assisted in revising and reviewing the manuscript.

\section{Acknowledgments}
 This project was supported by the National Institute of Dental and Craniofacial Research (R03DE027486), the National Institute of General Medical Sciences (R01GM126257). D.C. was supported by the Predoctoral Training Grant (T32GM135117) provided by the National Institute of General Medical Sciences. B.A.C. was supported by the National Institute of Environmental Health Sciences (P30ES000002). The Pediatric HIV/AIDS Cohort Study (PHACS) network was supported by the Eunice Kennedy Shriver National Institute of Child Health \& Human Development (NICHD) and other NIH institutes through grants to the Harvard T.H. Chan School of Public Health (P01HD103133 and HD052102) and with Tulane University School of Medicine (HD052104). The Health Professionals Follow-Up study (U01 CA 167552) and Nurse's Health Study (U01 CA176726) are supported by the National Cancer Institute. The content of this manuscript is solely the responsibility of the authors and does not necessarily represent the official views of the National Institutes of Health

 \section{Data availability}
 The data used in this study cannot be made publicly available for confidentiality reasons. To request data access, follow the instructions provided on the \textcolor{blue}{\href{https://phacsstudy.org/Our-Research/Data-and-Specimen-Sharing}{Pediatric HIV/AIDS Cohort Study}}, \textcolor{blue}{\href{https://www.nurseshealthstudy.org/researchers}{Nurse's Health Study}}, or \textcolor{blue}{\href{https://www.hsph.harvard.edu/hpfs/for-collaborators/}{Health Professionals Follow-up Study}} websites. 

\bibliography{sn-bibliography}

\end{document}